\documentclass[]{aastex631}
\usepackage{multirow}

\definecolor{red}{RGB}{250,0,0}

\shorttitle{ICL in TNG}
\shortauthors{Yoo et al.}

\graphicspath{{./}{figures/}}

\begin{document}

\title{Tracing Dark Matter in the Central Regions of Galaxy Clusters Using Galaxies, Gas, and Intracluster Light in TNG300: Connections to Cluster Dynamical State}

\correspondingauthor{Jihye Shin}
\email{jhshin@kasi.re.kr}

\author[0000-0002-6841-8329]{Jaewon Yoo}
\affiliation{Quantum Universe Center, Korea Institute for Advanced Study, 85 Hoegi-ro, Dongdaemun-gu, Seoul 02455, Korea}

\author[0000-0001-5135-1693]{Jihye Shin}
\affiliation{Korea Astronomy and Space Science Institute (KASI), Daedeokdae-ro, Daejeon 34055, Korea}
\affiliation{University of Science and Technology (UST), Gajeong-ro, Daejeon 34113, Korea}

\author[0000-0003-3428-7612]{Ho Seong Hwang}
\affiliation{Astronomy Program, Department of Physics and Astronomy, Seoul National University, Seoul 08826, Korea}
\affiliation{SNU Astronomy Research Center, Seoul National University, 1 Gwanak-ro, Gwanak-gu, Seoul 08826, Republic of Korea}
\affiliation{Australian Astronomical Optics - Macquarie University, 105 Delhi Road, North Ryde, NSW 2113, Australia}

\author[0000-0002-5513-5303]{Cristiano G. Sabiu}
\affiliation{Natural Science Research Institute (NSRI), University of Seoul, Seoul 02504, Korea}

\author[0000-0003-4032-8572]{Hyowon Kim}
\affiliation{Departamento de Física, Universidad Técnica Federico Santa María, Av. España 1680, Valparaíso, Chile}

\author[0000-0002-9434-5936]{Jongwan Ko}
\affiliation{Korea Astronomy and Space Science Institute (KASI), Daedeokdae-ro, Daejeon 34055, Korea}
\affiliation{University of Science and Technology (UST), Gajeong-ro, Daejeon 34113, Korea}

\author[0000-0003-0283-8352]{Jong Chul Lee}
\affiliation{Korea Astronomy and Space Science Institute (KASI), Daedeokdae-ro, Daejeon 34055, Korea}

\begin{abstract}
Recent studies have highlighted the potential of intracluster light (ICL) as a dark matter tracer. Moreover, the ICL co-evolves with the brightest cluster galaxy (BCG) and the host cluster, making it a valuable tool for understanding cluster dynamics.
In this study, we utilize 426 galaxy clusters (with total mass $M_{\rm tot}>10^{14} M_{\odot}$ at $z=0$) simulated in the cosmological hydrodynamical simulation Illustris TNG300 to compare the spatial distributions of dark matter, member galaxies, gas, and ICL and assess their effectiveness as dark matter tracers in the central regions of clusters at $R_{\rm vir}<0.3$.
We apply the Weighted Overlap Coefficient (WOC), a methodology for quantifying the similarity of two-dimensional spatial distributions, to various components of the galaxy clusters at different dynamical stages.
Our findings reveal that the spatial distributions of both ICL combined with the BCG and gas closely resemble the dark matter distribution, with higher fidelity observed in more relaxed galaxy clusters with earlier half-mass epochs.
These results demonstrate that the BCG+ICL component serves as an effective tracer of dark matter, consistent with previous observational studies linking cluster light to mass. Moreover, the degree of spatial similarity between the BCG+ICL and dark matter distributions appears to reflect the dynamical state of the cluster.

\end{abstract}


\section{Introduction} \label{sec:intro}
Galaxy clusters are the largest and most massive gravitationally bound systems in the Universe, comprising vast reservoirs of dark matter, hot diffuse gas, and a diverse population of galaxies. By virtue of their scale, they serve as natural laboratories for studying the hierarchical growth of structure, the interplay between baryonic and dark matter, and the dynamics driving galaxy evolution \citep[e.g.][]{2009ApJ...699.1595P}. In particular, the distribution of different mass components within clusters encodes key information about their assembly histories and dynamical states. Understanding these distributions—and the processes that shape them—provides a window into how clusters build up their mass over cosmic time, offering crucial insights into the broader picture of structure formation in a cosmological context.

One key aspect of galaxy clusters that reflects their evolutionary state is the phenomenon of \textit{intracluster light (ICL)}—stars bound to the gravitational potential of the galaxy cluster but not associated with any individual galaxy. The ICL is believed to form through various dynamic interactions, such as galaxy-galaxy encounters or between galaxies and the cluster's potential. As a result, ICL has emerged as a powerful tool for uncovering the assembly history and dynamical evolution of galaxy clusters \citep{2019A&A...622A.183J, 2021ApJ...910...45M, 2021MNRAS.508.2634Y, 2021Galax...9...60C}. 
In addition, ICL behaves as a collisionless component, much like dark matter, responding primarily to the cluster’s global gravitational potential.  
This similarity suggests that the spatial distribution of ICL should mirror that of dark matter. As such, ICL offers a unique observational window into both the dynamical state and the underlying dark matter structure of galaxy clusters. Leveraging this dual role, recent studies have demonstrated the potential of ICL as a \textit{luminous tracer of dark matter}, reinforcing its value in both cosmological and astrophysical contexts \citep{2019MNRAS.482.2838M, 2020MNRAS.494.1859A, 2022ApJS..261...28Y, 2023A&A...679A.159D,2024ApJ...965..145Y}.

In our previous work \citep{2022ApJ...934...43S}, we explored how well galaxies, gas, and ICL trace the spatial distribution of dark matter in and around 426 galaxy clusters from the Illustris TNG300 simulation \citep{2018MNRAS.475..648P, 2018MNRAS.475..676S, 2019ComAC...6....2N}. To quantify this, we performed an elliptical fitting to the density contours of each component and examined shape parameters at multiple radial distances. Among the three density maps—galaxies (mass-weighted number density), gas, and brightest cluster galaxy (BCG)+ICL—the galaxy map exhibited the strongest agreement with the ellipticity of dark matter across all radii (0.5, 1, and 2 $R_{\mathrm{vir}}$).
Additionally, we assessed the dynamical state of the clusters by computing their virial ratios. We found that more virialized clusters exhibit a stronger spatial alignment between dark matter and other baryonic components, whereas dynamically disturbed systems show weaker correspondence. However, as noted in \cite{2022ApJ...934...43S}, the elliptical fitting approach has limitations: It may not capture fine-scale features in inner regions and does not account for disconnected substructures. This motivated us to reexamine the analysis using a nonparametric comparison method.

To evaluate the similarity between two-dimensional spatial distributions, we developed the \textit{Weighted Overlap Coefficient (WOC)} method \citep{2022ApJS..261...28Y, 2024ApJ...965..145Y}, a novel approach designed to compare distributions without assuming a predefined relationship between their signal strengths. We argue that conventional methods may introduce biases when such assumptions are required and propose WOC as a more effective alternative for assessing morphological similarities in spatial fields.
The WOC method quantifies the overlap fraction between two distributions at multiple density threshold levels while incorporating signal strength weighting and normalizing the similarity metric within a 0 to 1 scale. This nonparametric approach is both intuitive and robust, avoiding numerical errors typically associated with contour fitting and enabling reliable similarity measurements across various smoothing scales, central alignments, and binning strategies.
Furthermore, WOC is particularly effective in cases involving disconnected structures or masked regions containing multiple significant substructures, such as the Bullet and Coma clusters, which are undergoing active merging processes. 

In \cite{2024ApJ...965..145Y}, we applied the WOC method to the hydrodynamical cosmological simulation Horizon Run 5 \citep{2021ApJ...908...11L}, demonstrating that the distribution of gas and BCG+ICL effectively traces the dark matter distribution. We also found that their spatial similarity to dark matter correlates with a galaxy cluster's dynamical state. However, the Horizon Run 5 simulation concludes at a redshift of approximately 0.625, limiting its applicability for comparisons with deep-imaging observational data in the local universe.
To bridge this gap and explore trends in the nearby universe, we need to extend this analysis to lower redshifts. Additionally, the Horizon Run 5 study did not incorporate virial ratio calculations—one of the most theoretically motivated measures of a cluster’s dynamical state. In contrast, our previous study based on Illustris TNG300 \citep{2022ApJ...934...43S} included virial ratio measurements, providing a valuable framework for assessing cluster relaxation.
Illustris TNG300 includes a large sample of galaxy clusters, 426 at redshift zero, which makes it a promising dataset for future work to compare simulation-based findings with observations of the local universe.

This paper has the following structure.
In \S2, we define the simulated data we will use and introduce our method.
We describe the cluster components and the dynamical state parameters used in the analysis in \S3. We present our results and discuss them in \S4. We conclude and summarize them in \S5.

\section{Data and Method} \label{sec:method}
\subsection{Simulation Data} \label{subsec:data}
For this study, we use the same data set as \cite{2022ApJ...934...43S}, which is publicly available data from TNG300 \citep{2018MNRAS.475..648P, 2018MNRAS.475..676S, 2019ComAC...6....2N}. The TNG300 is part of the Illustris TNG project \citep{2018MNRAS.473.4077P}, a successor to the original Illustris simulation \citep{2014Natur.509..177V}. The Illustris TNG suite includes three simulation volumes with side lengths of approximately 300, 100, and 50 Mpc: TNG300, TNG100, and TNG50. Among them, TNG300 offers the largest volume, making it ideal for studying galaxy clusters with minimal cosmic variance, even though it has the lowest dark matter resolution at $5.9\times10^7 M_{\odot}$ (1.48 kpc).

Since our primary interest lies in examining the large-scale morphology of cluster mass distributions rather than small-scale structures that depend on resolution, we choose TNG300 for its superior statistical representation. Additionally, the projected mass density maps used in this analysis are smoothed using a Gaussian kernel with $\sigma$ values between 90 and 240kpc, ensuring that the spatial resolution of TNG300 does not affect our results.

Our galaxy cluster sample is selected from the Friends-of-Friends (FoF) halo catalog at $z=0$, where halos are identified based on a linking length of $b=0.2$. We define galaxy clusters as FoF halos with masses above $10^{14} M_{\odot}$, which is a commonly used lower limit for clusters. Note that this study does not cover galaxy groups below this threshold. Our final selection includes 426 clusters, spanning a mass range of $10^{14} < M_{\mathrm{vir}}/M_{\odot} < 2.9\times10^{15}$ and sizes between $0.9<R_{\mathrm{vir}}/\mathrm{Mpc}<2.4$. In our analysis, we consider halos and subhalos in simulations as counterparts to galaxy clusters and cluster galaxies in real observations.

\subsection{Weighted Overlap Coefficient} \label{subsec:woc}
The similarity of the two spatial distributions is quantified as the WOC \citep{2022ApJS..261...28Y}, which is a value between 0 and 1.
To calculate the similarity between two components, we follow the steps below. 
\begin{itemize}
    \item First, we prepare two-dimensional maps of the two components smoothed at the same angular or spatial scale. 
    \item Then, we determine the signal center in the reference map by selecting the pixel with the maximum surface density and measure the one-dimensional radial profile. While different choices of the signal center can slightly alter the contour shapes and, consequently, the WOC result, this effect has been shown to be minor \citep{2022ApJS..261...28Y}.
    \item From each of the requested radii values, we identify the corresponding signal level (pixel value) in the one-dimensional radial profile.
    \item Next, for each of the signal levels identify the contour level in the reference map and compute their areas.
    \item We also find the contour levels in the comparison map, starting from the highest pixel value and gradually lowering the threshold until each contour matches the area of the reference map. This step allows us to quantify the overlapping area as a percentage. Importantly, the WOC method does not require normalization between different units in reference and comparison maps, since it depends only on the relative signal strength within each map. 
    \item Finally, we measure the area of overlap at each signal level and use the values to parametrize the similarity of the two maps of different components. We do this using $n$ overlapping fractions multiplied by weightings, including the inverse of the area fraction of each level to the lowest level, the signal level at the contour in the reference component, and the signal level at the contour in the comparison component.
\end{itemize}
Suppose we compare a reference map $A$ with a comparison map $B$ using $n$ sets of matched regions $A_i$ and $B_i$ with ${\rm area}(A_i)={\rm area}(B_i)$. Let the $i=1$ level correspond to the highest level. 
The WOC is then defined as
\begin{equation}
\mathrm{WOC}(A,B)=\frac{ \sum\limits_{i=1}^{n} f_{i} \left(w_{i}+w_{\rho_A,i}+w_{\rho_B,i}\right)}{\sum\limits_{i=1}^{n} 
\left(w_{i}+w_{\rho_A,i}+w_{\rho_B,i}\right)},
\label{eq:woc}
\end{equation}
where 
\begin{itemize}
    \item $f_{i}={\rm area}(A_i \bigcap B_i) /{\rm area}(A_i)$ represents the fractional overlap between the $i$-th contour regions of the two maps.
    \item $w_i={\rm area}(A_i)^{-1}/\sum_j {\rm area}(A_j)^{-1}$ is a normalized weight that assigns greater importance to higher-level contour areas.
    \item $w_{\rho_A,i}=\rho_{A,i}/\sum_j \rho_{A,j}$ and $w_{\rho_B,i}=\rho_{B,i}/\sum_j \rho_{B,j}$ are additional weights that prioritize higher-density threshold values.
    \item $\rho_{A,i}$ and $\rho_{B,i}$ denote the density threshold at $i^{th}$ level for maps $A$ and $B$, respectively.
\end{itemize}

This formulation ensures that the WOC effectively captures both spatial overlap and the relative significance of different contour levels between two maps (e.g., dark matter and ICL) based on area and density. Our primary focus is on assessing the spatial correspondence between the two maps, rather than their relative signal strengths or the specific shapes of their profiles.
The WOC method is particularly advantageous for handling disconnected regions and does not require the computation of individual contours, making it less susceptible to bias when applied to masked maps.
For a detailed discussion of the methodology, robustness tests, and comparisons with alternative methods, see \cite{2022ApJS..261...28Y}. 

\section{Analysis} \label{sec:analysis}
\subsection{Galaxy Cluster Components} \label{subsec:component}
\begin{figure}
\centering
\includegraphics[width=0.95\textwidth,trim={3cm 1cm 3cm 1.5cm},clip]{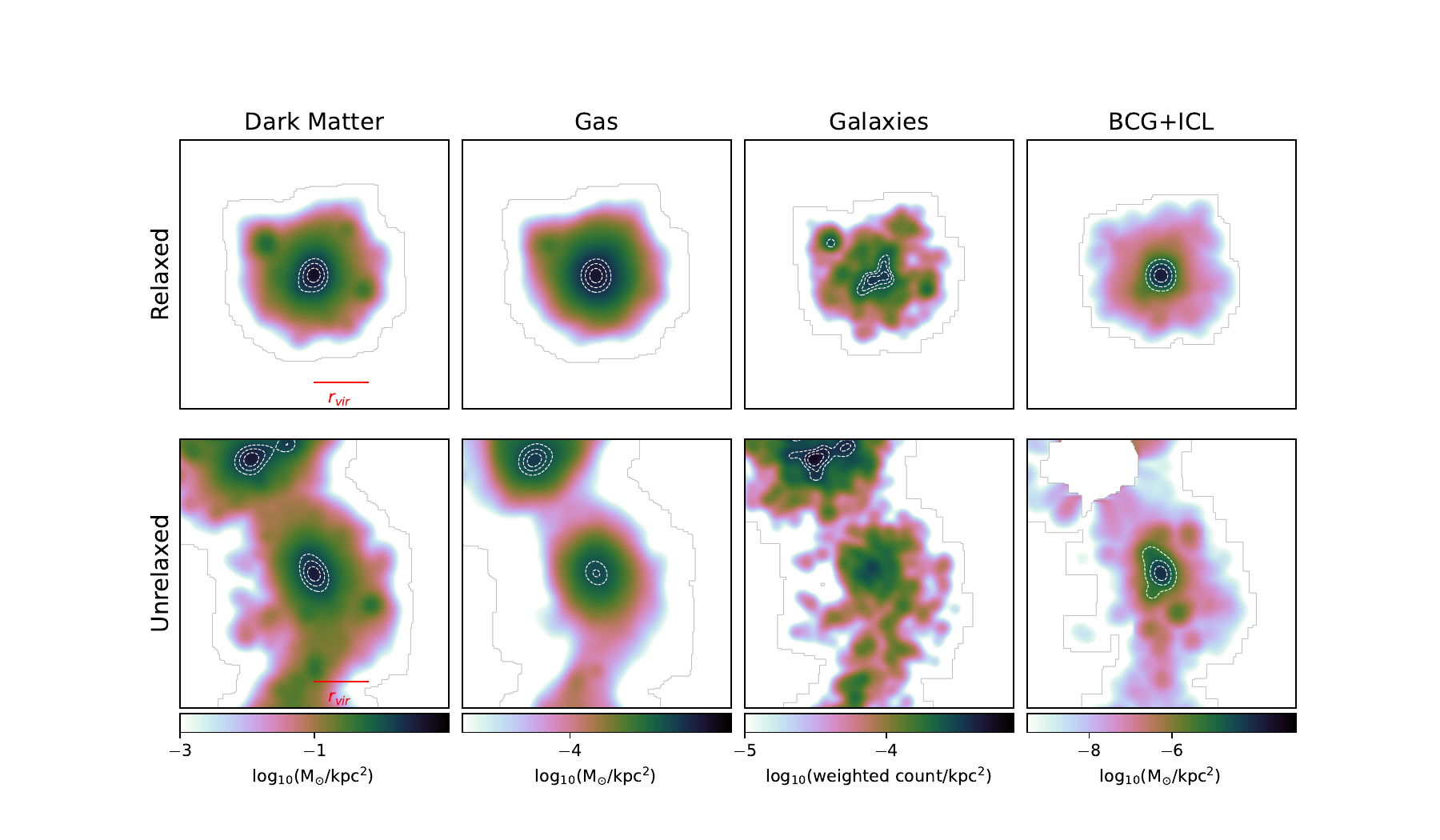}
\caption{Galaxy clusters in the TNG300 simulation. The upper row shows an example of an early-formed ($z_{m/2}\geq 0.71$; relaxed) galaxy cluster, and the lower row shows a late-formed ($z_{m/2}\leq 0.455$; unrelaxed) galaxy cluster. 
The scales of images are 5 $r_{\text{vir}} \times$ 5 $r_{\text{vir}}$ of the cluster. For each galaxy cluster, the dark matter, gas, galaxies, and BCG+ICL components are shown from the first column to the fourth column, respectively (see details in Section \ref{subsec:component}). Dashed lines indicate the density levels corresponding to the azimuthally averaged radial density profile at 0.1, 0.2, and 0.3 virial radii, which define the regions where the WOC—our metric for quantifying spatial similarity between components—is computed.
\label{fig:example2}
}
\end{figure}

To investigate which component best traces the spatial distribution of dark matter in galaxy clusters, we analyze three key baryonic elements: gas, galaxies, and ICL. Specifically, we consider (i) the gas surface density map, (ii) the galaxy mass-weighted number density map, and (iii) the combined BCG+ICL stellar surface density map. These components are derived from the same dataset used in \citet{2022ApJ...934...43S}, based on the IllustrisTNG300. Building on these elements, we apply the WOC method to quantify the spatial similarity between dark matter and each component, aiming to identify the most accurate dark matter tracer among them.

Using the snapshot at $z=0$, we generate projected mass density maps for dark matter, gas, and stars for each galaxy cluster. To achieve this, we extract all particles and cells associated with each cluster and project them along three different planes: $x-y$, $x-z$, and $y-z$. While this approach provides multiple independent projections per cluster, exploring a broader range of projection angles in future work could further enhance the effective sample size. The spatial distribution of these particles and cells is then binned into a two-dimensional array of $900\times900$ pixels$^2$ using the count-in-cell (CIC) algorithm available in the IDL astronomy user's library. 
The CIC algorithm divides a spatial domain into a regular grid of cells and counts the number of particles—or sums their mass—within each cell to construct a discrete density field. In simulation data, particle positions are provided as continuous coordinates in three-dimensional space. To create a projected density map, the simulation volume is projected onto a two-dimensional plane and divided into a grid. The CIC method is then applied to compute the surface density in each cell, resulting in a two-dimensional array that can be used to generate a surface density map.
Each map covers an area extending $\pm3R_{\mathrm{vir}}$ from the cluster center, which corresponds to the region of minimum gravitational potential energy. In these maps, 15 pixels correspond to a physical scale of 0.1$R_{\mathrm{vir}}$.

{\bf Gas:} 
Traditionally, X-ray observations of hot gas have been used to trace the total mass of galaxy clusters, as the gas is confined within their gravitational potential wells \citep{2001Natur.409...39B}. In this study, the gas component of galaxy clusters is shown in the second column of Figure \ref{fig:example2}.
Note that, we do not distinguish between cold and hot gas phases. The majority of the gas is hot, except in the central regions of galaxies. However, this has a minimal effect on our analysis due to the relatively coarse resolution of our projected density maps (30–80 kpc).

{\bf Galaxies:}
Observational studies based on dense redshift surveys suggest that galaxy number density can serve as a useful proxy for estimating the overall mass distribution of galaxy clusters \citep{2014ApJ...797..106H}. Furthermore, our previous simulation study \citep{2022ApJ...934...43S} indicates that the mass-weighted number density of galaxies traces the dark matter distribution even more effectively than either galaxy number density or galaxy velocity dispersion.
To construct the galaxy mass-weighted number density map, we extract the positions and stellar masses of cluster member galaxies from the subhalo catalog. Our sample includes galaxies with masses down to approximately $1.2\times10^9M_{\odot}$, which corresponds to the minimum mass required for identification by the SUBFIND algorithm \citep{2001MNRAS.328..726S} (20 dark matter particles). 
The SUBFIND algorithm identifies gravitationally self-bound substructures (subhalos) within larger dark matter halos (such as FoF halos) by locating locally overdense regions and removing unbound particles through an iterative unbinding process.
The mass-weighted number density is computed on a two-dimensional grid of $300\times300$ pixels$^2$, spanning $\pm3R_{\mathrm{vir}}$ from the cluster center. Since the number of galaxies is significantly lower than that of dark matter, gas, and stellar particles, we reduce the pixel resolution by a factor of three compared to other density maps ($900\times900$ pixels$^2$) to improve the statistical significance of galaxy counts. The resulting galaxy maps are then smoothed using a Gaussian kernel with $\sigma=5$ pixels, ensuring consistency with the smoothing scale used for other density maps.
Because cluster member galaxies, treated as test particles, have a discrete distribution and a broad mass range ($10^9-10^{12}M_{\odot}$), a simple linear mass-weighting approach would introduce artificial clumpiness in the density maps. To mitigate this effect, we rescale the mass weights so that galaxies at the lower and upper mass limits have weights of one and ten, respectively (excluding the BCG).

{\bf BCG+ICL:} 
The BCG resides at the deepest gravitational potential well of the galaxy cluster, while the ICL is expected to follow the overall gravitational potential of the cluster. As a result, the combined BCG+ICL distribution serves as a promising tracer of dark matter \citep{2019MNRAS.482.2838M, 2020MNRAS.494.1859A, 2022ApJS..261...28Y, 2024ApJ...965..145Y}.
To focus on the global distribution of diffuse ICL, we remove local stellar density peaks associated with individual galaxies, as these stars primarily trace the mass distribution of their host galaxies rather than the overall cluster potential. Individual galaxies are therefore treated as separate components, referred to as ``Galaxies'', which includes the BCG as well. We identify them using the SUBFIND subhalo catalog \citep{2001MNRAS.328..726S} and mask regions extending up to 2$R_h$ from each galaxy center, where $R_h$ is the half-mass radius of the galaxy. However, small galaxies with $2R_h$ smaller than half a pixel width are excluded from masking and are instead considered part of the diffuse ICL population. Additionally, we do not apply a mask to the BCG, as this could remove a significant portion of the cluster’s central region.
To analyze two-dimensional distributions, we smooth the projected stellar mass density maps using a Gaussian kernel with a standard deviation of 15 pixels via the GAUSS\_SMOOTH/IDL routine. For stellar mass density maps, masked regions are treated as missing data when updating unmasked pixel values during smoothing. While most masked areas are effectively interpolated using surrounding pixels, large masked regions (radius $>3\sigma$) can introduce artifacts such as distortions or holes, as pixels within a $3\sigma$ radius may lack sufficient real data for interpolation. However, since these masks are located around satellite galaxies, where the BCG+ICL surface density is rather low, such artifacts do not influence the contours of interest and thus do not affect the WOC results. 
Ultimately, the smoothed stellar mass density map after masking represents the diffuse ICL distribution, including the BCG.
The fourth column in Figure \ref{fig:example2} illustrates the distribution of BCG+ICL within the galaxy clusters analyzed in this study.

To ensure a fair comparison between different components and maintain equal numerical precision in the WOC measurement process, we rescale the density maps of dark matter, gas, and BCG+ICL to match the $300\times300$ pixel$^2$ resolution used for the galaxy maps.

We note that the resulting WOC value tends to increase with larger smoothing kernel sizes, as low-density or empty regions become progressively filled in, leading to greater apparent overlap between maps. However, this increase eventually plateaus—beyond a certain kernel size, the WOC value converges \citep{2022ApJS..261...28Y}. This behavior underscores the importance of applying a sufficiently large and consistent smoothing kernel across all maps being compared, which is particularly crucial for future applications of the WOC method to observational data.

\subsection{Dynamical State Indicators} \label{subsec:dynamic}
\begin{figure}
\centering
\includegraphics[width=0.9\textwidth,trim={3cm 0 3cm 1cm},clip]{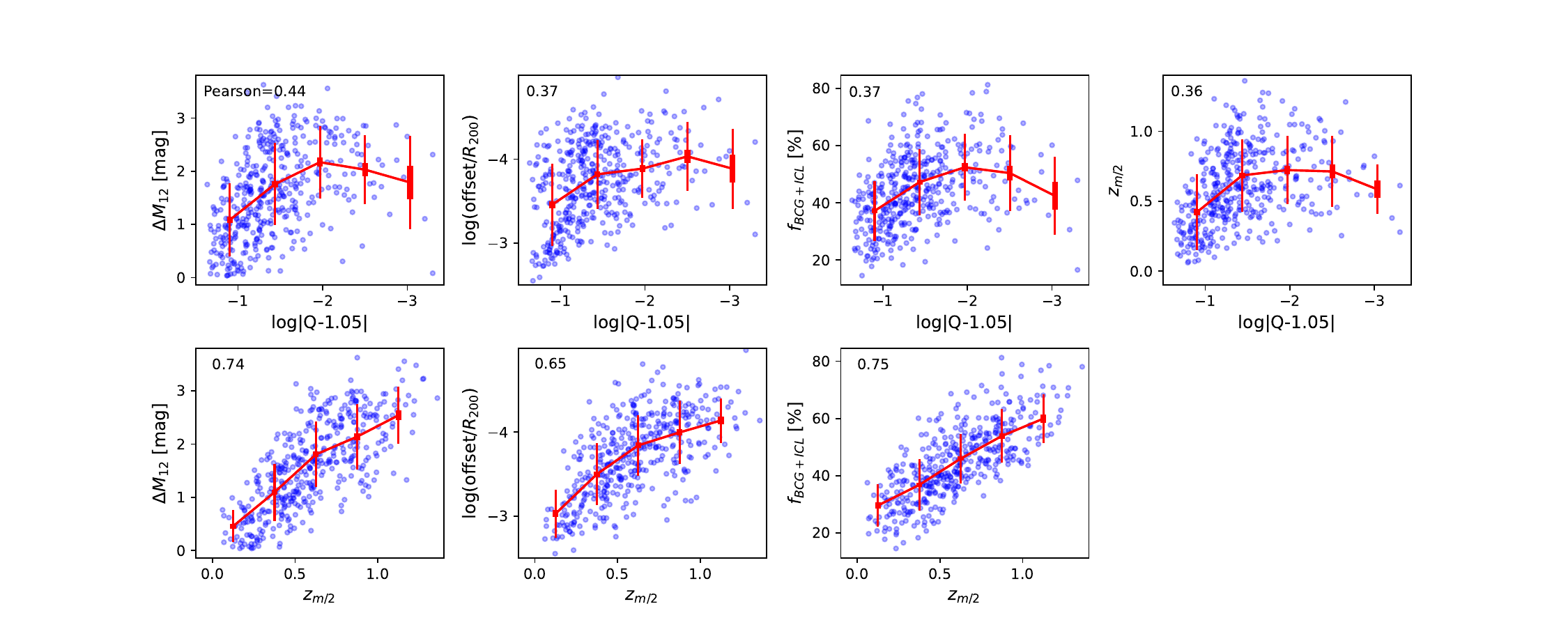}
\caption{Simulation-based and observationally accessible dynamical state parameters used in this study and their relation to the other parameters. (Top) The magnitude gap between BCG and the second brightest galaxy, BCG-cluster center offset, BCG+ICL fraction, and half-mass epoch for different virial ratios are plotted from left to right, respectively. (Bottom) Similarly, from left to right, we show the magnitude gap between BCG and the second brightest galaxy, BCG-cluster center offset, and BCG+ICL fraction for different half-mass epochs. 
For the virial ratio and offset parameters, we take the absolute deviation from the standard value (which represents the most relaxed state) or normalize by the cluster's virial radius, followed by a logarithmic transformation to visually enhance the representation of the relaxation state.
Thin vertical lines are standard deviations, and thick vertical lines are standard errors calculated as the standard deviation divided by the square root of the sample size. For every sub-plot, increasing the $x$-axis and increasing the $y$-axis should indicate a more relaxed status. Pearson coefficients are indicated in the upper left corner of each subplot.}
\label{fig:dyn}
\end{figure}

The dynamical state of a galaxy cluster plays a crucial role in determining how well its baryonic components trace the underlying dark matter distribution. To investigate this, we employ several well-established indicators of cluster relaxation, each offering unique insights into the system's evolutionary history and current state. These include simulation-based metrics such as the virial ratio and half-mass epoch, as well as observationally accessible indicators like the magnitude gap, center-of-mass offset, and BCG+ICL fraction. The relationships among these indicators are illustrated in Figure \ref{fig:dyn}, highlighting the extent to which observable proxies reflect the dynamical state inferred from simulations.
We quantify the strength of the correlation using the Pearson coefficient, where values closer to 1 indicate a strong positive correlation, values near 0 suggest no correlation, and values approaching -1 signify a strong anti-correlation. 

{\bf Virial Ratio ($Q$):}
The virial ratio is a fundamental measure of a cluster's dynamical equilibrium. For an idealized, isolated system in virial equilibrium, the virial theorem predicts $2K+U=0$, where $K$ is the total kinetic energy and $U$ is the gravitational potential energy. However, galaxy clusters are not isolated systems; they experience ongoing accretion and interactions with their surroundings. To account for this, we include a surface pressure term $E_s$ in our calculation of the virial ratio:
\begin{equation}
Q = \frac{2K - E_s}{|U|}.
\label{eq:virial}
\end{equation}
Here, $E_s$ corrects for the energy contribution from infalling material at the cluster boundary \citep{1961hhs}. Taking into account the surface pressure term and the distribution of virial ratios within our galaxy cluster sample (see Fig. 12 in \cite{2022ApJ...934...43S}), we classify clusters with virial ratios close to 1.05 as dynamically relaxed. Deviations from this benchmark indicate progressively higher levels of dynamical disturbance. For a detailed discussion of the virial ratio and its calculation, see \cite{2022ApJ...934...43S}.

{\bf Half-mass Epoch ($z_{m/2}$):} 
In the $\Lambda$CDM framework, galaxy clusters grow hierarchically through mergers and accretion. The redshift at which a cluster assembles half of its final mass ($z_{m/2}$) provides a useful proxy for its dynamical state. Clusters with early formation times ($z_{m/2}\gg0$) have had sufficient time to relax and virialize, whereas those with recent major mergers ($z_{m/2}\approx0$) are likely to exhibit disturbed morphologies and ongoing dynamical activity. This parameter is particularly valuable for distinguishing between relaxed and unrelaxed systems in simulations \citep{1996MNRAS.281..716C, 2021A&A...651A..56G}.

This is clearly illustrated in Figure \ref{fig:example2}, which contrasts an early-formed, relaxed cluster (upper row) with a late-formed, unrelaxed system (lower row). The relaxed cluster shows smooth, symmetric distributions of dark matter, gas, and BCG+ICL, consistent with a long period of dynamical equilibrium. In contrast, the unrelaxed cluster exhibits elongated and scattered structures across all components, indicative of recent major merging activity. These morphological differences are directly linked to the clusters' formation histories, with the relaxed system having assembled most of its mass at higher redshifts ($z_{m/2}>1$) and the unrelaxed system undergoing significant mass growth at lower redshifts ($z_{m/2}<0.5$).

These two simulation-based parameters—virial ratio and half-mass epoch—are not directly measurable in observational data, as they require access to three-dimensional particle velocities or detailed halo mass assembly histories across multiple simulation snapshots, which are only available in simulations. In the context of future observational applications, it is therefore essential to develop a ``translator" that connects these simulation-derived proxies to observable quantities. Figure \ref{fig:dyn} provides such a connection by illustrating the correlation between simulation-based parameters (x-axis) and observationally accessible indicators (y-axis), offering a pathway to interpret observable features in terms of underlying cluster dynamics.

{\bf Magnitude Gap ($\Delta M_{12}$):}
The magnitude gap between the BCG and the second brightest galaxy ($\Delta M_{12}$) is a widely used observational indicator of cluster relaxation. A large gap ($\Delta M_{12}>2$ in the $r$--band) suggests that the cluster has experienced efficient early accretion, leading to the formation of a dominant BCG and minimal recent mergers. Such systems, often referred to as fossil clusters, are typically relaxed and exhibit extended X-ray emission \citep{2003MNRAS.343..627J}. In contrast, clusters with small magnitude gaps are more likely to be dynamically active. In the bottom left panel of Figure \ref{fig:dyn}, we find that $\Delta M_{12}$ shows a strong correlation with $z_{m/2}$, with a Pearson correlation coefficient of 0.74. This suggests that the magnitude gap, derived from just the two brightest galaxies in a cluster, can serve as an effective proxy for the half-mass epoch, which reflects the entire mass assembly history of the galaxy cluster.

We note that TNG and other hydrodynamic simulations are known to produce overgrown BCGs due to limitations in modeling AGN feedback. Consequently, the fraction of clusters with large magnitude gaps (e.g., $\Delta M_{12}>2$) is typically higher in simulations than in observations, where such systems are observed in only about 4–10\% of clusters. Although we do not explicitly correct for this discrepancy, our analysis focuses on the relative trends between the magnitude gap and other dynamical indicators rather than the absolute values. Therefore, this limitation does not significantly affect our conclusions.

{\bf Center-of-Mass Offset:} 
The spatial offset between the cluster's center of mass and the BCG provides another measure of dynamical state. In relaxed systems, the BCG is expected to reside at the bottom of the gravitational potential well, coinciding with the center of mass. However, in merging or disturbed clusters, this alignment is disrupted, leading to measurable offsets. We quantify this offset as the distance between the BCG and the cluster's center of mass, normalized by the virial radius $R_{vir}$ \citep{2017MNRAS.464.2502C, 2021A&A...651A..56G}.
In Figure \ref{fig:dyn}, we see that the spatial offset also exhibits a correlation with the half-mass epoch, obtaining a Pearson coefficient of 0.65.

{\bf BCG+ICL fraction ($f_{\rm BCG+ICL}$):} 
The fraction of stellar mass contained in the BCG and ICL is a powerful indicator of cluster relaxation. Both the BCG and ICL are enriched by dynamical processes such as tidal stripping and galaxy mergers, which are more prevalent in relaxed systems. The BCG+ICL fraction has been extensively studied in both simulations and observations, with higher fractions typically associated with older, more relaxed clusters \citep{2014MNRAS.437.3787C, 2019ApJ...871...24C}. Notably, this parameter exhibits a strong correlation with the magnitude gap $\Delta M_{12}$, making it a robust and complementary indicator of dynamical state.
In the bottom panel of Figure \ref{fig:dyn}, we plot the BCG+ICL fraction against the half-mass epoch and find a highly correlated behavior reaching a Pearson coefficient of 0.75.

Figure \ref{fig:dyn} illustrates the relationships between these dynamical state indicators. 
While the virial ratio shows only weak correlations with other indicators (upper panels), the half-mass epoch ($z_{m/2}$) exhibits a strong correlation with minimal scatter when compared to observable indicators such as the magnitude gap, center-of-mass offset, and BCG+ICL fraction (lower panels).
The fourth column of Figure \ref{fig:dyn} further demonstrates the weak relationship between the two simulation-based indicators, the virial ratio and the half-mass epoch. 

If we have a reliable proxy for representing the dynamical state of a system, it should exhibit a strong correlation with other established proxies that characterize the system's dynamical state. Based on this reasoning, our results suggest that the half-mass epoch serves as a more robust simulation-based indicator compared to the virial ratio. As also discussed in the Appendix \ref{sec:appendix1}, the virial ratio shows somewhat distinct behavior from the other indicators, potentially because it captures information from the outer regions of the cluster, while other indicators—such as the magnitude gap, offset, and BCG+ICL fraction—are more sensitive to the central region. This difference in spatial sensitivity may explain the weaker correlation between the virial ratio and WOC values measured at small cluster-centric radii (0.1, 0.2, and 0.3 virial radii).
We will further discuss the feasibility of using the virial ratio as a dynamical state indicator in Appendix \ref{sec:appendix1}.

Among the observable indicators, the BCG+ICL fraction exhibits the strongest correlation with the half-mass epoch, followed by the magnitude gap, with the center-of-mass offset showing the weakest correlation. Although the BCG+ICL fraction and the magnitude gap are independent measures (as the BCG+ICL fraction does not include information about the second brightest galaxy), both indicators show strong correlations with the half-mass epoch, which reflects the system’s mass accretion history. This suggests that the degree to which the BCG dominates the stellar component—captured by both indicators—is an important metric for representing the dynamical state of the cluster.

Additionally, we examined another observable indicator, Kuiper's V-statistic \citep{KUIPER196038,2023MNRAS.525.4685S}, which quantifies the asymmetry in the spatial distribution of member galaxies within a cluster. However, we did not find a significant correlation between this statistic and other dynamical indicators in our dataset.

{\bf Combination of the Indicators}: 
Previous studies suggest combining several dynamical state indicators to estimate the dynamical state of galaxy clusters in a more robust way \citep{2020MNRAS.492.6074H, 2021A&A...651A..56G, 2024ApJ...970..165K}.
In particular, \citet{2024ApJ...970..165K} demonstrated that the three observable indicators considered in this study—magnitude gap, center-of-mass offset, and satellite stellar mass fraction (equivalently, the BCG+ICL fraction)—serve as effective proxies when used in combination to characterize the relaxation state of galaxy clusters.

\begin{figure}
\centering
\includegraphics[width=0.9\textwidth,trim={0.1cm 0 0.1cm 0.1cm},clip]{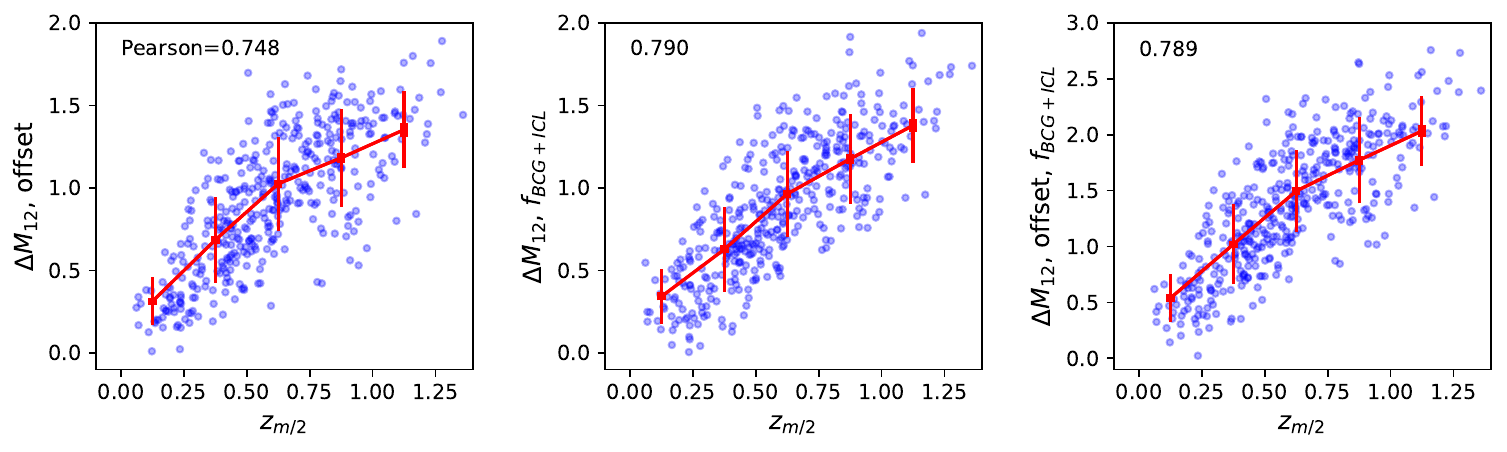}
\caption{Combination of dynamical state parameters and the relation with the half-mass epoch. (Left) The magnitude gap between BCG and the second brightest galaxy, and the BCG-cluster center offset, are combined, and their correlation with the half-mass epoch is plotted. (Middle) The magnitude gap between BCG and the second brightest galaxy and the BCG+ICL fraction are combined, and their correlation with the half-mass epoch is plotted. (Right) The magnitude gap between BCG and the second brightest galaxy, the BCG-cluster center offset, and the BCG+ICL fraction are combined, and the relation for different half-mass epochs is plotted. Thin vertical lines are standard deviations, and thick vertical lines are standard errors. Pearson coefficients are indicated in the upper left corner of each subplot.}
\label{fig:dyn_comb}
\end{figure}

In this work, we take a simplified approach to combining dynamical state indicators. Each indicator is normalized to span a common range, with 0 representing the most dynamically disturbed (unrelaxed) state and 1 representing the most relaxed state. The normalized indicators are then summed to produce a composite relaxation score for each galaxy cluster.
In the leftmost panel of Figure \ref{fig:dyn_comb}, we see the relation between the combination of the magnitude gap and the center-of-mass offset, and the half-mass epoch. It exhibits a stronger correlation than either indicator alone, with a Pearson coefficient of 0.748.
We also consider the combination of the magnitude gap and the BCG+ICL fraction in the middle panel of Figure \ref{fig:dyn_comb}. Their correlation with the half-mass epoch is increased with a Pearson coefficient of 0.790, better than either alone.
We next combine the three indicators together, as can be seen in the right panel of Figure \ref{fig:dyn_comb}. Their combination also exhibits a high correlation with the half-mass epoch, with a Pearson coefficient of 0.789. However, the Pearson coefficient is comparable to the middle panel, suggesting that the center-of-mass offset is not adding much to the correlation. These results demonstrate that combining multiple observable indicators—particularly the magnitude gap and BCG+ICL fraction—provides a more robust proxy for cluster relaxation than any single indicator alone, with diminishing returns when adding the center-of-mass offset.

\section{Results and Discussions} \label{sec:result}

We analyzed the spatial distribution similarity between different components in galaxy clusters to determine which component best traces dark matter and how this varies with the dynamical state of the clusters. The components considered include (1) all gas particles, (2) the mass-weighted number density of galaxies, encompassing both the BCG and satellite galaxies, and (3) stellar particles associated with the BCG and ICL.

The binning for measuring the WOC was set at 0.1, 0.2, and 0.3 virial radii ($R_{200}$) for each cluster. This choice aligns with observational studies of ICL, where detection is typically limited to scales of approximately 300 kpc due to background fluctuations. Additionally, we adopted the same binning scheme as our previous study using the WOC method \citep{2024ApJ...965..145Y}, facilitating direct comparison between different simulations. While we also performed the analysis at larger scales (0.1, 0.3, and 0.5 virial radii, or 0.5 virial radius only) to directly compare with one of the scales examined in \citet[][see Appendix \ref{sec:appendix2}]{2022ApJ...934...43S}, we find that the reliability of galaxies as dark matter tracers improves significantly at larger scales, particularly when measured at 0.5 virial radius only, highlighting a clear scale dependence in their performance.

Dark matter always serves as the reference map, while gas, galaxies, and BCG + ICL maps are used for comparison. During the radial profile measurement in the WOC calculation, the center is defined as the dark matter density peak. However, it is important to note that the WOC method itself does not require a predefined center when comparing maps, as it independently determines contour density levels based on equal-area regions for each map.

\subsection{Dark Matter Tracers}
\label{subsec:result1}

\begin{table}[t]
	\centering
 	\caption{Summary of WOC Results. The measurement scale for the WOC was set at 0.1, 0.2, and 0.3 virial radii for each cluster. Among the 426 clusters, we categorize 142 early-formed clusters ($z_{m/2} \geq 0.71$) as relaxed, 141 late-formed clusters ($0 < z_{m/2} \leq 0.455$) as unrelaxed, and the remaining clusters as middle, ensuring that each group contains a similar number of samples. There is one galaxy cluster where the half-mass epoch ($z_{m/2}$) is not measured, which is excluded from the grouping. The WOC results of each group are calculated by taking the mean of the WOC results of the individual clusters, where all three projections ($x-y$, $x-z$, and $y-z$) of WOC were measured. Standard deviations for each group are shown in parentheses. The mean standard deviations of the WOC result of three different projections are indicated below the WOC value. }
	\begin{tabular}{c|cccc}
            \hline & All 426 clusters & Relaxed 142 clusters & Middle 142 clusters & Unrelaxed 141 clusters \\
            \hline WOC (DM, Gas) & 0.859 ($\pm$ 0.104) & 0.925 ($\pm$ 0.044) & 0.866 ($\pm$ 0.082) & 0.784 ($\pm$ 0.117) \\
            std ($xy, yz, zx$) & 0.035 & 0.018 & 0.031 & 0.056\\
            \hline WOC (DM, Galaxy) & 0.487 ($\pm$ 0.169) &  0.493 ($\pm$ 0.166)& 0.497 ($\pm$ 0.156) & 0.470 ($\pm$ 0.183) \\
            std ($xy, yz, zx$) &  0.119 & 0.129 & 0.124 & 0.103\\
            \hline WOC (DM, BCG+ICL) & 0.875 ($\pm$ 0.099)& 0.925 ($\pm$ 0.047) & 0.881 ($\pm$ 0.077)& 0.818 ($\pm$ 0.125)\\
            std ($xy, yz, zx$) & 0.046 & 0.027 & 0.047 & 0.065 \\
            \hline 
	\end{tabular}

	\label{tab:table2} 
\end{table}

We calculated the WOC between dark matter and other cluster components to determine which component's spatial distribution most closely matches that of dark matter. 
The measured WOC results for all 426 galaxy clusters are summarized in Table \ref{tab:table2}.
We take a mean of three different projection angles ($x-y$, $x-z$, and $y-z$) for each cluster and take a mean and standard deviation of all the cluster results. The mean values of the standard deviation of the three projection angles was also computed (see std ($xy, yz, zx$) in Table \ref{tab:table2}). 

We find that the BCG+ICL component closely traces the spatial distribution of dark matter. Our results align with a broad body of observational studies examining the correlation between light and mass in galaxy clusters, including work from the lensing community (e.g., \citealt{2005ApJ...621...53B, 2009ApJ...703L.132Z}, and subsequent studies), as well as more recent efforts using globular clusters to trace the ICL and, by extension, the underlying mass distribution (e.g., \citealt{2022ApJ...940L..19L, 2023A&A...679A.159D}).

The WOC value between dark matter and galaxies is significantly lower (0.487) within the measured radial range (0.1, 0.2, and 0.3 virial radii) compared to other candidate components used to trace dark matter.
In contrast, both BCG+ICL and gas show much stronger correlations with dark matter, with WOC values of 0.875 and 0.859, respectively.
For different projection angles ($x-y$, $x-z$, and $y-z$), WOC (DM, galaxy) exhibits a larger deviation (0.119), whereas WOC (DM, gas) and WOC (DM, BCG+ICL) show smaller deviations of 0.035 and 0.046, respectively. This suggests that the spatial distribution of galaxies is more sensitive to projection effects compared to gas and BCG+ICL, which maintain more consistent alignment with dark matter across different viewing angles.

These findings are consistent with previous results from the GRT simulation \citep{2022ApJS..261...28Y} and Horizon Run 5 \citep{2024ApJ...965..145Y}, both of which demonstrated that BCG+ICL more faithfully traces the underlying dark matter distribution than satellite galaxies. While some of the anisotropies observed in the galaxy distribution may stem from infalling structures along cosmic filaments, such anisotropies are generally more pronounced in the outer regions of clusters. In the inner regions, galaxy orbits are expected to become more isotropic (e.g., \citealt{2004A&A...424..779B}). A more plausible explanation for the weaker correlation between galaxies and dark matter in the cluster cores is that galaxies are biased tracers—fewer galaxies are found in the inner regions, both due to the actual number density and due to limitations in simulation-based galaxy identification schemes, which may struggle to distinguish individual galaxies near the BCG. Additionally, projection effects along the line of sight can further obscure the true spatial distribution of galaxies, making them less reliable tracers of the dark matter potential. In contrast, the BCG, typically an elliptical galaxy with a smooth extended envelope, and the ICL, characterized by high velocity dispersion and a diffuse spatial distribution, provide a more isotropic and smoother representation of the cluster’s central potential. Moreover, the dark matter’s gravitational potential itself is inherently smoother than the localized density peaks of stellar particles associated with satellite galaxies.

However, this result contradicts our previous study using the same dataset \citep{2022ApJ...934...43S}, where the ellipticity of the member galaxy distribution was found to best match that of dark matter, rather than the ellipticity of gas or BCG+ICL. In \cite{2022ApJ...934...43S}, ellipticity was measured at 0.5, 1, and 2 virial radii, whereas in this study, WOC is measured at 0.1, 0.2, and 0.3 virial radii.
This discrepancy suggests that at larger scales—specifically in the outer regions and beyond the galaxy cluster—the broad distribution of galaxies effectively represents the global extent of dark matter. In contrast, BCG+ICL dominates in the central region of the cluster, where its detailed shape more accurately reflects the morphology of dark matter. 

To ensure a common measurement scale with \citet{2022ApJ...934...43S}, we also computed the WOC at 0.1, 0.3, and 0.5 virial radii, as well as at 0.5 virial radius alone, as described in Appendix \ref{sec:appendix2}. The results obtained at 0.1, 0.3, and 0.5 virial radii were consistent with those measured at 0.1, 0.2, and 0.3 virial radii, reinforcing the robustness of our findings. Notably, the WOC value for galaxies at 0.5 virial radius alone is comparable to that of BCG+ICL and gas, highlighting the potential of galaxies to serve as reliable tracers of dark matter at larger scales—an insight that may be particularly relevant for wide-field imaging surveys with limited surface brightness sensitivity.

\subsection{Relation with Dynamical State}
\label{subsec:result2}

\begin{figure}
\centering
\includegraphics[width=1.0\textwidth,trim={4cm 1cm 4cm 1cm},clip]{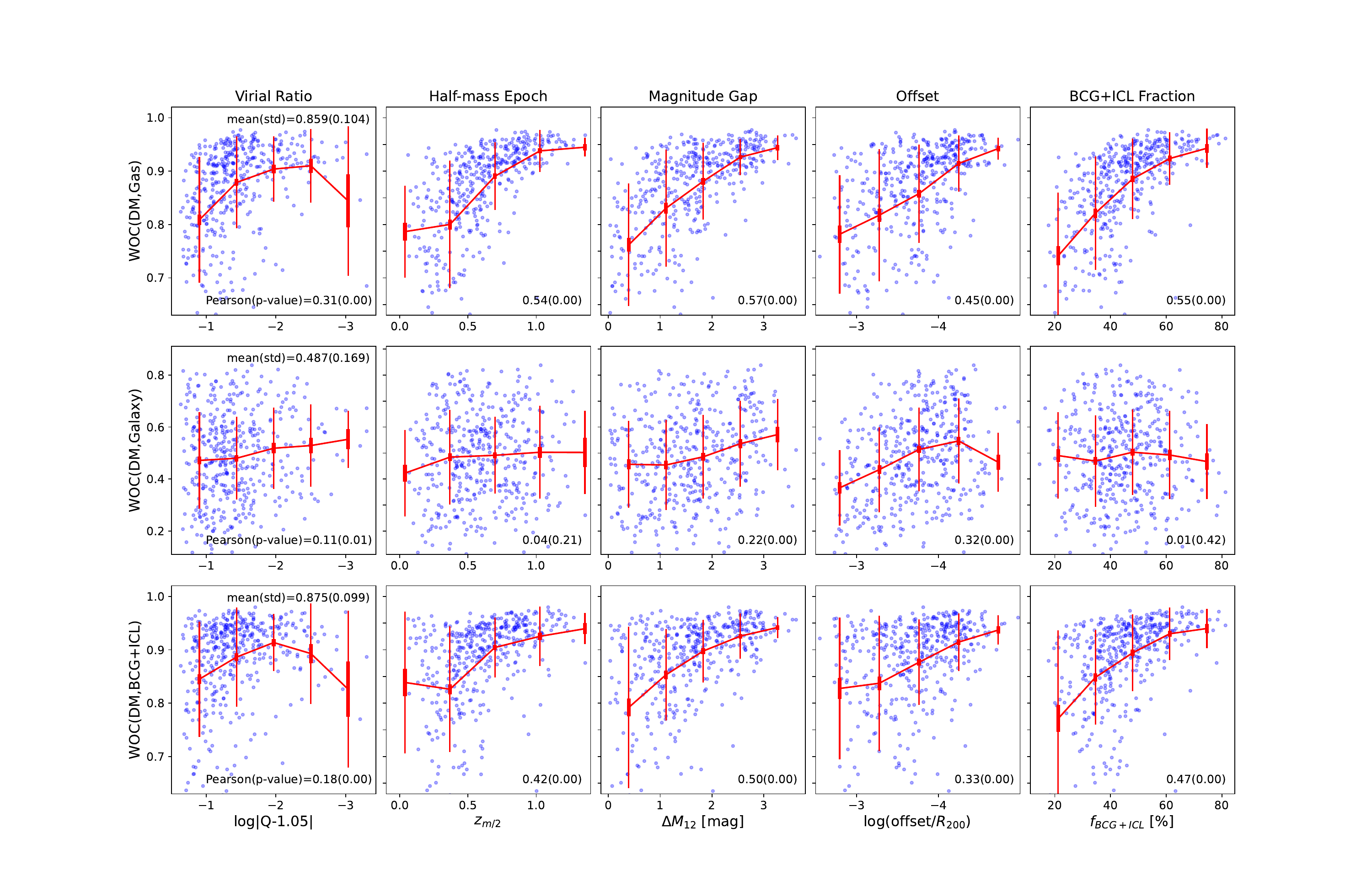}
\caption{WOC results for various components and dynamical state indicators. Spatial distribution similarity between dark matter and gas particles (first row), all galaxies (second row), and BCG+ICL (third row) is plotted for various relaxedness proxies. For every subplot, increasing the $x$-axis represents a more relaxed status. Blue dots are the WOC results for individual galaxy clusters. Red trend lines represent the median value of WOC in each bin, thin vertical red lines are standard deviations, and thick vertical red lines are standard errors. The medians are calculated from the full distribution, including values outside the plotted range. Pearson coefficients and p-values are indicated in the lower right corner of each subplot.}
\label{fig:woc_dyn123}
\end{figure}

Over time, as galaxy clusters undergo virialization, the spatial distributions of dark matter and other cluster components tend to become more aligned. If a clear relationship exists between a cluster's dynamical relaxation and the similarity between dark matter and its tracers, this similarity could, in turn, serve as an indicator of the cluster’s dynamical stage.
To investigate this, we analyze the correlation between the WOC of dark matter and different components—gas, galaxies, and BCG+ICL—alongside various dynamical state indicators introduced in Section \ref{subsec:dynamic}. These indicators include the virial ratio, half-mass epoch ($z_{m/2}$), magnitude gap ($\Delta M_{12}$), BCG-cluster center offset, and BCG+ICL fraction (see Figure \ref{fig:woc_dyn123}).

Figure \ref{fig:woc_dyn123} illustrates the relationship between WOC and cluster relaxedness. The $x$-axis represents increasing dynamical relaxation, while the $y$-axis indicates greater spatial similarity between dark matter and the respective component. Each blue dot corresponds to an individual galaxy cluster, while red lines show the median WOC trend. Error bars denote standard deviation (thin vertical red lines) and standard error (thick vertical red lines).
As in Figure \ref{fig:dyn}, the strength of these correlations is measured using the Pearson coefficient. Additionally, we assess statistical significance using the p-value, with smaller values indicating stronger evidence of correlation.

In relaxed galaxy clusters, the spatial distribution of dark matter aligns more closely with BCG+ICL or gas compared to dynamically young clusters. This suggests that these components serve as more reliable tracers of dark matter in well-virialized clusters (first and third rows in Figure \ref{fig:woc_dyn123}). In contrast, the similarity between galaxies and dark matter, measured by WOC(DM, Galaxy), does not show a clear trend with increasing cluster relaxation (second row in Figure \ref{fig:woc_dyn123}).

Among the dynamical state indicators, the virial ratio (first column) exhibits the weakest correlation, with the lowest Pearson coefficient. As discussed in \ref{subsec:dynamic}, this suggests that the virial ratio is a less reliable measure of a cluster’s dynamical state compared to the half-mass epoch. A more detailed assessment of its feasibility as a dynamical state proxy is provided in Appendix \ref{sec:appendix1}. On the other hand, the correlation between WOC and the half-mass epoch ($z_{m/2}$) reveals a clear trend for both BCG+ICL and gas (second column in Figure \ref{fig:woc_dyn123}).

The strongest correlations for WOC(DM, Gas) and WOC(DM, BCG+ICL) are observed with the magnitude gap, with Pearson coefficients of 0.57 and 0.50, respectively. The second highest correlation is found with the BCG+ICL fraction, yielding Pearson coefficients of 0.55 for gas and 0.47 for BCG+ICL.
The magnitude gap and BCG+ICL fraction serve as optically observable indicators of a galaxy cluster’s dynamical state. 
Our findings suggest that in clusters with a relatively high BCG+ICL fraction (60$\sim$80\%), the spatial distribution of BCG+ICL or gas can trace dark matter with an accuracy of approximately 90$\sim$95\%.  

However, a major challenge in utilizing this property lies in the difficulty of observing ICL due to its low surface brightness. Measuring ICL remains a notoriously complex task, as current telescopes are limited by systematics, and different data analysis methods adopt varying definitions of ICL, leading to inconsistencies in results. Systematic errors from optical components and high background fluctuations further complicate the detection of the faint diffuse light from ICL. Given the sensitivity of results to the chosen ICL definition, it is essential to adopt a consistent methodology and apply it across a large dataset. Establishing a standardized detection and analysis approach for ICL across numerous galaxy clusters is crucial. The upcoming LSST survey from the Rubin Observatory is expected to play a pivotal role in advancing ICL studies as a potential dark matter tracer and a cluster's dynamical state indicator.

\section{Conclusions} \label{sec:conclusion}

We have investigated how well various baryonic components trace the underlying dark matter distribution in 426 massive galaxy clusters ($M_{\rm tot}>10^{14}M_{\odot}$) from the Illustris TNG300 simulation at $z=0$. We used the Weighted Overlap Coefficient (WOC) method to measure the spatial similarity between dark matter and several components—gas, mass-weighted galaxies, and the combination of the BCG and ICL, focusing on the central regions of clusters at $R_{\rm vir}<0.3$.
By quantifying the overlap across multiple smoothed density levels, the WOC approach remains robust to morphological complexity, disconnected substructures, and potential masking of bright galaxies.

Our WOC analysis demonstrates that the BCG+ICL component in Illustris TNG300 closely reproduces the spatial morphology of the dark matter. Our results thereby bolster the classic `light traces mass' paradigm in cluster cores, in agreement with both gravitational lensing analyses and novel globular cluster studies. This convergence of simulation and observational evidence strengthens the case for using the BCG+ICL distribution as a reliable proxy for dark matter in galaxy clusters, especially for well-relaxed systems.

Across all clusters, we find that the spatial distribution of BCG+ICL or gas aligns significantly better with dark matter than does the distribution of galaxies alone. The mean WOC values between dark matter and BCG+ICL (or dark matter and gas) both exceed $\sim 0.85$, while the dark matter-galaxy similarity is substantially lower ($\sim 0.49$). The diffuse nature of the ICL and the extended, relatively isotropic gas component both reflect the deeper gravitational potential more closely than the sparser satellite system. As demonstrated in Appendix \ref{sec:appendix2}, galaxies more effectively trace dark matter at larger scales, with notably improved performance when measured at 0.5 virial radius alone, underscoring the scale dependence of their reliability.

We compared the WOC values to multiple indicators of dynamical relaxation including the simulation-based indicators such as the virial ratio and the half-mass epoch ($z_{m/2}$), and observable indicators such as the magnitude gap $\Delta M_{12}$, the center-of-mass offset, and the fraction of stellar mass in BCG+ICL.
Between the two simulation-based indicators, the half-mass epoch exhibits a significantly stronger relationship with other observable indicators than the virial ratio.
For both gas and BCG+ICL, WOC values rise systematically as clusters appear more relaxed. In clusters with large $\Delta M_{12}$ (and thus a dominant BCG), the dark matter-BCG+ICL similarity can reach $\sim90\%$. The correlation between WOC and cluster relaxation state is relatively weak when using the galaxy distribution as a tracer.

Among observationally accessible indicators, we confirm that a high BCG+ICL fraction ($f_{\rm BCG+ICL}$) correlates strongly with early cluster half-mass epoch and high dark matter–ICL spatial similarity. 
This highlights the potential of ICL-based tracers for probing dark matter distributions and cluster dynamical states in next-generation low surface brightness imaging surveys, provided that deep data reduction methods can robustly isolate diffuse intracluster starlight.

These findings offer a consistent picture of how baryonic matter—in its diffuse stellar and gaseous forms—converges toward the underlying dark matter distribution in mature clusters, while satellite galaxies exhibit more substructure. Future work will compare these simulation-based results with ongoing and upcoming deep optical surveys that can better capture the faint outskirts of BCG+ICL. 
Unlike in simulations, dark matter is not directly observable in the real universe. Instead, cluster mass distributions are inferred through strong and weak gravitational lensing analyses, which introduce their own sources of systematic uncertainty. For instance, \cite{2025OJAp....8E..37P} highlights persistent divergences in mass reconstructions of galaxy clusters, even with an increasing number of lensing constraints. In this context, the ICL offers an alternative and independent tracer of dark matter, with fewer—or at least different—systematic uncertainties. Based on the findings of this study, the BCG+ICL component holds promise as a practical dark matter tracer and offers a powerful avenue for improving our understanding of cluster assembly and relaxation across cosmic time.

Building on these insights, a natural extension of this study would be to adapt the methodology developed for ICL at the galaxy cluster scale to the galaxy scale. By analyzing diffuse tidal features around individual galaxies, we may be able to trace and characterize the spatial distribution of their dark matter halos. While this endeavor requires high-resolution hydrodynamical simulations, it aligns well with the advent of wide-field, deep, low-surface-brightness imaging from upcoming surveys, presenting exciting new opportunities to explore galaxy-scale dark matter structures.

In addition, the recently released TNG-Cluster simulation provides an excellent platform to further explore how various cluster properties—such as total mass, dynamical state, and filamentary connectivity—affect the morphology and alignment of diffuse components. Such comprehensive studies will deepen our understanding of the interplay between dark matter and baryons across different cosmic environments.

\begin{acknowledgments}
We would like to thank the anonymous referee whose careful and insightful comments allowed us to improve the presentation of this work significantly.
J.Y. was supported by a KIAS Individual Grant (QP089902) via the Quantum Universe Center at Korea Institute for Advanced Study.
J.H.S. acknowledges support from the National Research Foundation of Korea grants (No. RS-2025-00516904, No. RS-2022-NR068800) funded by the Ministry of Science, ICT \& Future Planning. 
HSH acknowledges the support of the National Research Foundation of Korea (NRF) grant funded by the Korea government (MSIT), NRF-2021R1A2C1094577, Samsung Electronic Co., Ltd. (Project Number IO220811-01945-01), and Hyunsong Educational \& Cultural Foundation.
C.G.S is supported via the Basic Science Research Program from the National Research Foundation of South Korea (NRF) funded by the Ministry of Education (2018R1A6A1A06024977) and by the NRF (RS-2025-00515276). HK acknowledges the support of the Agencia Nacional de Investigación y Desarrollo (ANID) ALMA grant funded by the Chilean government, ANID-ALMA-31230016.

\end{acknowledgments}

\vspace{5mm}
\facilities{Illustris TNG}

\software{astropy \citep{2013A&A...558A..33A,2018AJ....156..123A}
          }

\appendix
\restartappendixnumbering
\section{Virial Ratio and Dynamical State}
\label{sec:appendix1}

The virial ratio ($Q$) in this study is defined in Equation \ref{eq:virial}, where a system is considered virialized (relaxed) when the ratio approaches $Q\approx1$. At this value, the system has reached dynamical equilibrium, meaning that twice the kinetic energy approximately balances the potential energy. A virial ratio significantly above unity indicates excess kinetic energy and a dynamically disturbed state, while a lower value suggests a dominance of potential energy.

Considering the surface pressure term and the peak value of the histogram of virial ratios in our galaxy cluster sample \citep{2022ApJ...934...43S}, we classify clusters with a virial ratio near 1.05 as relaxed, while deviations from this value correspond to increasing levels of dynamical disturbance. We quantify this deviation as shown in Figure \ref{fig:dyn}.

Unlike other parameters, which show a one-directional trend, the virial ratio follows a roughly normal distribution, peaking at the most frequently occurring (relaxed) state. 
To assess this, without taking the absolute value or the logarithmic values, we examined trends for clusters with virial ratios and other dynamical state indicators, and checked if there are different trends for those clusters above and below the virial ratio $\sim$1.05, i.e., kinetic energy dominant case or potential energy dominant case.

\begin{figure}
\centering
\includegraphics[width=0.9\textwidth]{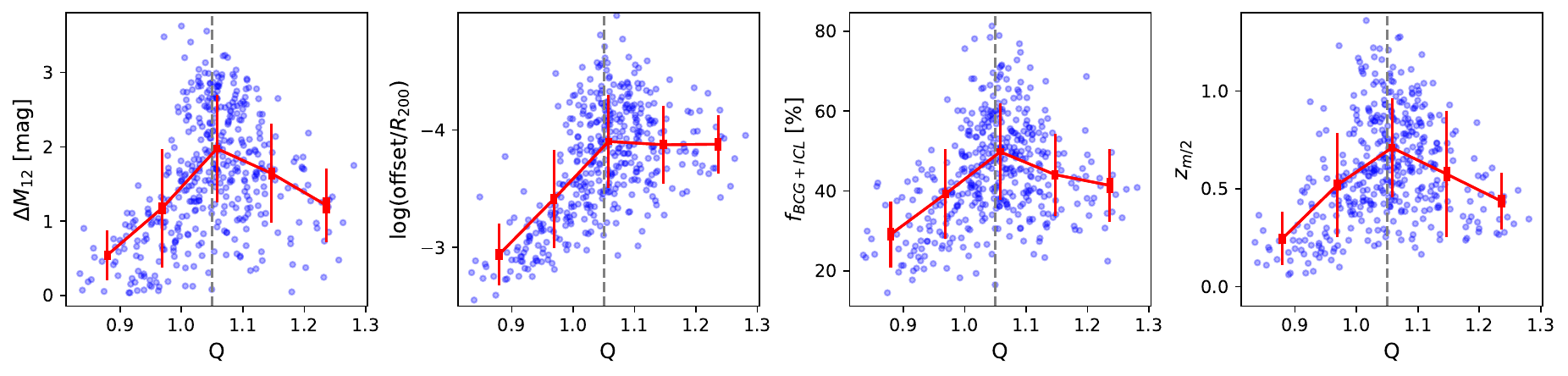}
\caption{Virial ratio ($Q$) and various dynamical state indicators. The magnitude gap between BCG and the second brightest galaxy, BCG-cluster center offset, BCG+ICL fraction, and half-mass epoch for different virial ratios are plotted from left to right, respectively. Blue dots show the results for individual galaxy clusters. For every subplot, increasing the $y$-axis represents a more relaxed status. Gray dashed vertical lines indicate where the virial ratio is 1.05. Red trend lines represent the median value in each bin, thin vertical red lines are standard deviations, and thick vertical red lines are standard errors.}
\label{fig:virial}
\end{figure}

As shown in Figure \ref{fig:virial}, except for the offset case (second panel), the distribution increases toward the peak at 1.05 and then declines. This reinforces the virial ratio as a reliable dynamical state proxy. However, the trend on the right side (higher virial ratios) appears somewhat weaker than on the left side. The predominance of the middle value also dampens the trend observed in Figure \ref{fig:dyn}, particularly around $\log \left| Q - 1.05 \right|$, where the most relaxed systems (right side of the $x$-axis, $\log \left| Q - 1.05 \right| \sim -3$) exhibit a weaker correlation.

One possible explanation for why the higher virial ratio clusters (higher relative kinetic energy) favor a smaller BCG-cluster center offset (plateau in second panel) could be that the high virial ratio clusters represent systems that are going through a merger with a first pass where the galaxies have high velocity but small potential energy.

\section{Larger Scale Measurement}
\label{sec:appendix2}

As in our previous work \citep{2024ApJ...965..145Y}, we first measured the WOC at 0.1, 0.2, and 0.3 virial radii, a radial range well matched to the observable extent of the ICL. To cross-validate our methodology against the ellipticity–based analysis of \citet{2022ApJ...934...43S}—which examined the same Illustris TNG300 dataset out to larger radii—we then repeated the WOC calculation at 0.1, 0.3, and 0.5 virial radii. Aligning the radial apertures in this way enables a direct comparison between the two independent approaches.

Attempting to measure WOC at 0.5, 1, and 2 virial radii—consistent with \cite{2022ApJ...934...43S}—is rather challenging. In these outer regions, the procedure often failed because, although the contour area was successfully defined in the reference map (dark matter), no corresponding area could be identified in the comparison maps (galaxies or BCG+ICL). This is because neither component sufficiently fills the space at larger radii to match the dark matter contour area: the galaxy distribution, while extended, is sparse and patchy, occupying a limited area, whereas the BCG+ICL component is smoothly distributed but strongly concentrated near the cluster center. Consequently, neither component adequately spans the spatial domain defined by the dark matter contours. From a physical standpoint, such cases of WOC non-detection indicate a significant mismatch in spatial distributions. While one could attempt to measure WOC by reversing the reference and comparison maps or applying extreme smoothing to artificially fill the missing regions, such approaches would undermine fair comparisons across components and distort the physical interpretation.

\begin{figure}
\centering
\includegraphics[width=1.0\textwidth,trim={4cm 1cm 4cm 1cm},clip]{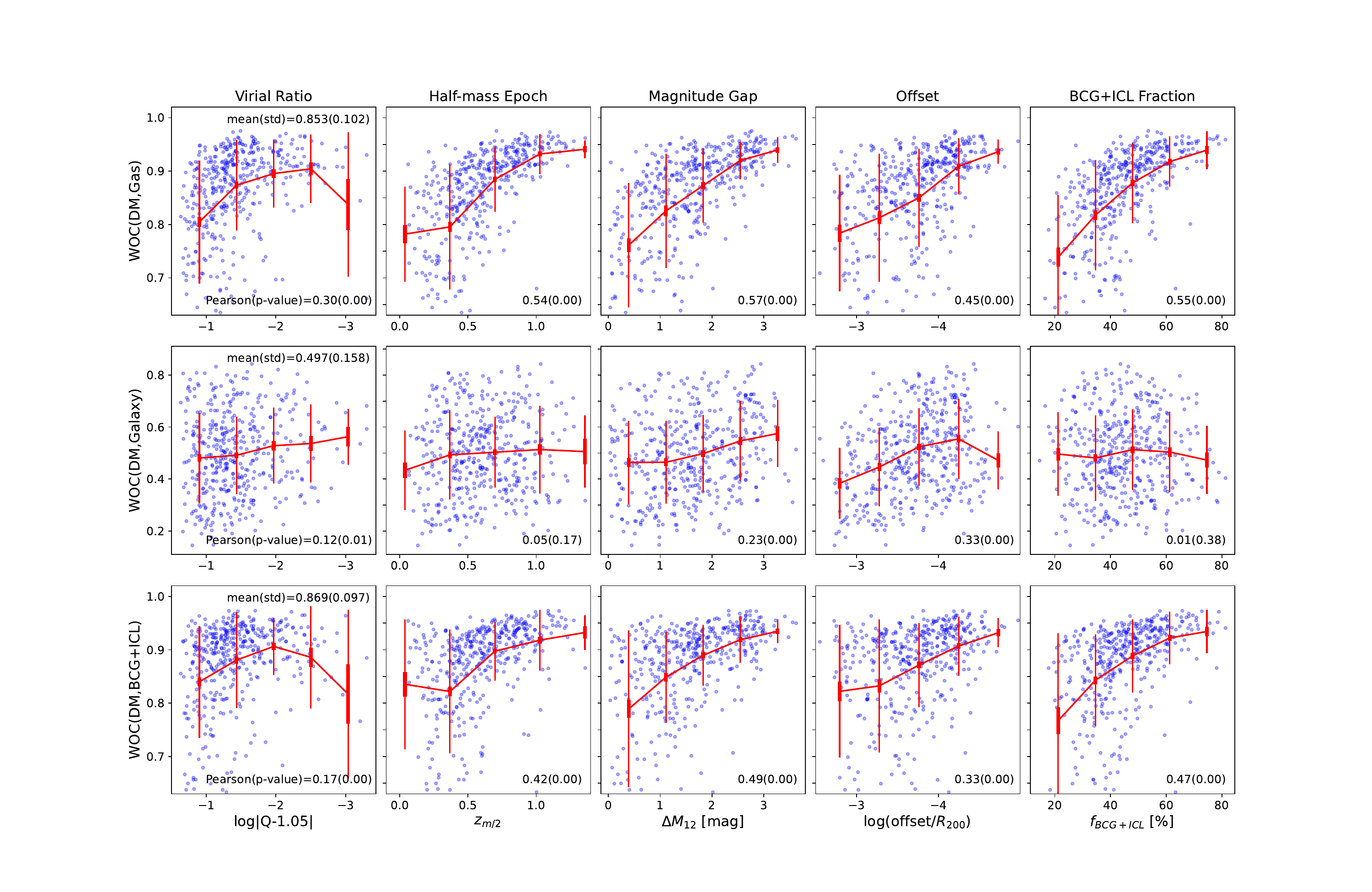}
\caption{WOC results for various components and dynamical state indicators. The WOC is calculated at 0.1, 0.3, and 0.5 virial radii. Spatial distribution similarity between dark matter and gas particles (first row), all galaxies (second row), and BCG+ICL (third row) is plotted for various relaxedness proxies. For every subplot, increasing the $x$-axis represents a more relaxed status. Blue dots are the WOC results for individual galaxy clusters. Red trend lines represent the median value of WOC in each bin, thin vertical red lines are standard deviations, and thick vertical red lines are standard errors. Pearson coefficients and p-values are indicated in the lower right corner of each subplot.}
\label{fig:woc_dyn135}
\end{figure}

To ensure at least one overlapping measurement scale, we instead calculated WOC at 0.1, 0.3, and 0.5 virial radii. As shown in Table \ref{tab:table3} and Figure \ref{fig:woc_dyn135}, the results and trends for 0.1, 0.3, and 0.5 virial radii align well with those obtained using 0.1, 0.2, and 0.3 virial radii (Table \ref{tab:table2} and Figure \ref{fig:woc_dyn123}). This consistency suggests that our findings are not limited to the central region of galaxy clusters at $R_{\rm vir}<0.3$ but remain robust across larger at $R_{\rm vir}<0.5$, observable scales.

The WOC method assigns greater weight to bins with stronger signals. As a result, when measuring WOC across multiple scales (0.1, 0.3, and 0.5 virial radii), the contribution from the innermost region (0.1 virial radius) is emphasized more than that from the outermost region (0.5 virial radius). To enable a more direct comparison with \cite{2022ApJ...934...43S}, which presented individual results at 0.5, 1, and 2 virial radii, we also performed WOC measurements using only the overlap area at 0.5 virial radius—that is, using a single radial bin. These results are presented in Table \ref{tab:table3}.

When using the combined binning at 0.1, 0.3, and 0.5 virial radii, BCG+ICL shows the best performance in tracing dark matter, followed by gas, with galaxies performing the poorest. However, for the single-scale measurement at 0.5 virial radius, gas becomes the most accurate tracer, followed by BCG+ICL, and then galaxies. Interestingly, the performance of galaxies improves significantly in this case, with the WOC value for galaxies increasing from 0.497 (in the multi-scale measurement) to 0.788 (in the single-scale measurement). This suggests that when the inner (0.1 virial radius) and intermediate (0.3 virial radius) regions are included, the overall alignment between galaxy distribution and dark matter falls below 50\%. In contrast, when considering only the 0.5 virial radius scale, galaxies can trace dark matter up to approximately 80\%. As discussed in Section \ref{subsec:result1}, this result supports the idea that the luminosity- or mass-weighted galaxy distribution serves as a more reliable tracer of dark matter at larger scales.

\begin{table}[t]
	\centering
 	\caption{Summary of WOC Results. The WOC was measured at 0.1, 0.3, and 0.5 virial radii, or at 0.5 virial radius only for each cluster. The dynamical state groupings of the sample clusters follow those defined in Table \ref{tab:table2}. For each group, the WOC values represent the mean across individual clusters, where measurements were performed in all three projections ($x-y$, $x-z$, and $y-z$). Standard deviations for each group are shown in parentheses. }
	\begin{tabular}{c|c|cccc}
            \hline & Binning [$R_{\rm vir}$] & All 426 clusters & Relaxed 142  & Middle 142  & Unrelaxed 141  \\ 
            \hline \multirow{2}{10.6em}{WOC (DM, Gas)} &0.1, 0.3, 0.5 & 0.853 ($\pm$ 0.102) & 0.920 ($\pm$ 0.042) & 0.859 ($\pm$ 0.080) & 0.781 ($\pm$ 0.116) \\
            &0.5 & 0.863 ($\pm$ 0.054) & 0.895 ($\pm$ 0.041) & 0.859 ($\pm$ 0.043) & 0.836 ($\pm$ 0.059)\\
            \hline \multirow{2}{10.6em}{WOC (DM, Galaxy)}&0.1, 0.3, 0.5  & 0.497 ($\pm$ 0.158) &  0.504 ($\pm$ 0.155)& 0.507 ($\pm$ 0.144) & 0.478 ($\pm$ 0.172) \\
             &0.5 &  0.788 ($\pm$ 0.072) & 0.788 ($\pm$ 0.065) & 0.800 ($\pm$ 0.060) & 0.778 ($\pm$ 0.088)\\
            \hline \multirow{2}{10.6em}{WOC (DM, BCG+ICL)}&0.1, 0.3, 0.5  & 0.869 ($\pm$ 0.097)& 0.918 ($\pm$ 0.047) & 0.874 ($\pm$ 0.076)& 0.815 ($\pm$ 0.122)\\
             &0.5 & 0.808 ($\pm$ 0.074) & 0.835 ($\pm$ 0.059) & 0.819 ($\pm$ 0.059) & 0.769 ($\pm$ 0.086) \\
            \hline 
	\end{tabular}

	\label{tab:table3} 
\end{table}


\bibliographystyle{aasjournal}

\end{document}